\documentclass[12pt]{article}
\usepackage{graphicx}

\newcommand\pubnumber{NuPhys2018-Kozhuharov}
\newcommand\pubdate{\today}

\def\Journal#1#2#3#4{{#1} {\bf #2}, #3 (#4)}

\def\PLB{{\em Phys. Lett.}  B}

\def\lnf{INFN - Laboratori Nazionali di Frascati, 
Via E. Fermi 40, 00044 Frascati, ITALY}
\def\sofia{Faculty of Physics, Sofia University, 5 J. Bourchier Blvd., 1164 Sofia, Bulgaria}
\def\support{\footnote{Present address: \sofia}}
\def\collaboration{\footnote{On behalf of the NA62 Collaboration:  R. Aliberti, F. Ambrosino, R. Ammendola, 
B. Angelucci, A. Antonelli, G. Anzivino, R. Arcidiacono, T. Bache,
M.  Barbanera,  J.  Bernhard
% ...
}}

\def\Title#1{\begin{center} {\Large #1 } \end{center}}
\def\Author#1{\begin{center}{ \sc #1} \end{center}}
\def\Address#1{\begin{center}{ \it #1} \end{center}}

\newcommand\pubblock{\rightline{\begin{tabular}{l} \pubnumber\\
         \pubdate  \end{tabular}}}
\newenvironment{Abstract}{\begin{quotation}  }{\end{quotation}}
\newenvironment{Presented}{\begin{quotation} \begin{center} 
             PRESENTED AT\end{center}\bigskip 
      \begin{center}\begin{large}}{\end{large}\end{center} \end{quotation}}

\def\beq{\begin{equation}}
\def\eeq#1{\label{#1}\end{equation}}
\def\eeqn{\end{equation}}

\def\beqa{\begin{eqnarray}}
\def\eeqa#1{\label{#1}\end{eqnarray}}
\def\eeqan{\end{eqnarray}}

\let\bar=\overbar

\def\Dslash{\not{\hbox{\kern-4pt $D$}}}
\def\dslash{\not{\hbox{\kern-2pt $\del$}}}

\def\msb{{\bar{\ssstyle M \kern -1pt S}}}

\begin{document}
\begin{titlepage}
\pubblock

\vfill
\Title{Search for heavy neutrinos at CERN SPS}
\vfill
\Author{ Venelin Kozhuharov\collaboration
\support
}
\Address{\lnf% \\
%\sofia
}
\vfill
\begin{Abstract}
%Among the different extensions of the Standard Model
%the neutrinos are the place where new phenomenology is 
%necessary and granted. 
The phenomenology in the neutrino sector 
%of the Standard Model
%necessarily 
requires physics beyond the Standard Model.
%new  phenomenology for the explanation of the observed 
%oscillation properties. 
One possibility is the existence of new massive 
leptonic states which could be probed at the high intensity machines. %, 
%like the SPS at CERN. 
The present results on heavy neutral leptons from the study of kaon decays in flight with 
the NA48/2 and NA62 experiments are presented 
and the future prospects for such searches 
at CERN SPS are discussed.

%Extension of the sensitivity reach 
%to heavy neutral leptons for given scenarios
%is possible through the exploitation
%of the beam dump technique and future 
%prospects are discussed. 

\end{Abstract}

\vfill
\begin{Presented}
 NuPhys2018, Prospects in Neutrino Physics
Cavendish Conference Centre, London, UK, December 19--21, 2018
\end{Presented}
\vfill
\end{titlepage}
\def\thefootnote{\fnsymbol{footnote}}
\setcounter{footnote}{0}

\section{Introduction}

A possible explanation of the neutrino oscillations 
is the extension of the Standard Model with three sterile Majorana neutrinos $N_i$ - 
the so called Neutrino Minimal Standard Model 
($\nu$MSM)\cite{bib:shaposhnikov}.
The mass of the lightest neutrino could be of the order of 1 keV/c$^2$, 
making it a suitable Dark matter candidate,
%and constitute the Dark Matter
while the mass of the others could be in the range from 100 MeV/c$^2$ to few GeV/c$^2$. 
The observed small masses of the left handed neutral leptons are provided 
through the see-saw 
mechanism.
The ($\nu$MSM) could also be extended by a scalar field $\chi$ to account 
for the inflation \cite{bib:tkachev-shap}.

%
%\begin{figure}[htb]
%\centering
%\includegraphics[width=0.5\textwidth]{neutrinos.png}
%\caption{Possible extension of the Standard Model with three neutral leptonic degrees of freedom.}
%\label{fig:neutrinos}
%\end{figure}

In meson decays, the heavy neutrinos are 
produced through mixing with the ordinary ones. 
For example, the rate for the decay $K^+ \to \ell^+ N$ is 
given by \cite{bib:nu-masses}

\begin{equation}
\Gamma (K^+ \to \ell^+ N ) =   \Gamma (K^+ \to \ell^+ \nu )  \times \rho_{\ell}(m_N) \times |U_{\ell 4}|^2,
\end{equation}
where $m_N$ is the heavy neutral lepton (HNL) mass, $|U_{\ell 4}|$ is the mixing parameter between the HNL and the 
neutrino corresponding to the lepton $\ell$, and 
$\rho_{\ell}(m_N)$ is a kinematic factor which includes 
also the helicity suppression  in the electron case. 
$\rho_{\ell}(m_N)$ is O(1) for most of the accessible $m_N$ range.

%leading to 
%\begin{equation}
%|U_{\ell 4}|^2 = \frac{Br (K^+ \to \ell^+ N )}{ Br (K^+ \to \ell^+ \nu ) \times	 \rho_{\ell}(m_N)}
%\end{equation}

The decay width of the HNL is proportional to 
$|U_{\ell 4}|^2\times m_{N}^3$. 
Depending on its mass and mixing, different scenarios are possible:
\begin{itemize}
 \item HNL decays within the fiducial region of the experiment. 
% only to SM final states. Then $\Gamma(N \to SM ) \sim |U_{\ell 4}|^2\times m_{N}^3$. 
 For $m_N < 500$ MeV/c$^2$ the possible final states are $N \to \pi^0 \nu$, $N \to \pi^{\pm}\mu^{\mp}$,
 $N \to \pi^{\pm} e^{\mp}$, $N \to \nu\nu\nu$. 
 
 \item If $|U_{\ell 4}|^2 < 10^{-4}$ then $\gamma c \tau_N >  $ 10 km and $N$ could be considered 
 invisible to the experimental apparatus. 
\end{itemize}

The presence of Majorana mass term could manifest itself in the 
Lepton Number Violating decay (LNV) $K^{\pm}\to\pi^{\mp}\mu^{\pm}\mu^{\pm}$.
%which might occur through the on-shell production of $N_i$. 
%The inflaton could be produced in $K^{\pm}\to\pi^{\pm}\chi$ decay, with 
%a subsequent decay to a muon pair. 
In addition, the existence of a heavy neutrino
with $2m_{\mu}< m_N < m_{K^\pm} - m_{\pi^\pm}$ might appear as a
resonance in the $M_{\pi\mu}$ mass spectrum of the 
Lepton Number Conserving (LNC) decay 
$K^{\pm}\to\pi^{\pm}\mu^{\pm}\mu^{\mp}$.

\section{Exclusive HNL searches}

A search for the decay $N \to \pi^{\pm}\mu^{\mp}$ of a HNL 
produced in $K^+ \to \ell^+ N$ has been performed 
by both NA48/2 and NA62 experiments. 

%\subsection{}
NA48/2 operated in 2003 and 2004 with
 simultaneous $K^+$ and $K^-$ beams with
(60 $\pm$ 3.7) GeV/c momentum. 
The kaon decay products were registered by the NA48 detector \cite{bib:na48}. 
A spectrometer consisting of four drift chambers separated by a dipole magnet
measured the momentum 
of the charged particles with resolution $\sigma(p)/p $ = (1.0  $\oplus $ 0.044  $p$ [GeV/c])\%. 
The timing and a fast trigger condition were provided by a scintillator hodoscope with time resolution of 150 ps.
The energy of photons and electrons was measured by a quasi-homogeneous liquid
krypton electromagnetic calorimeter, providing 
resolution $\sigma(E)/E = 3.2\%/ \sqrt{E} \oplus 9\%/ E \oplus  0.42\%$ (energy is in GeV). 
The charged particle identification was based on the ratio E/p. 
%It was also able to 
%provide charged particle identification based on the energy deposit by a particle with respect to 
%its momentum E/p.

\begin{figure}[hptb]
\minipage{0.5\textwidth}
\begin{center}
\includegraphics[width=0.9\textwidth]{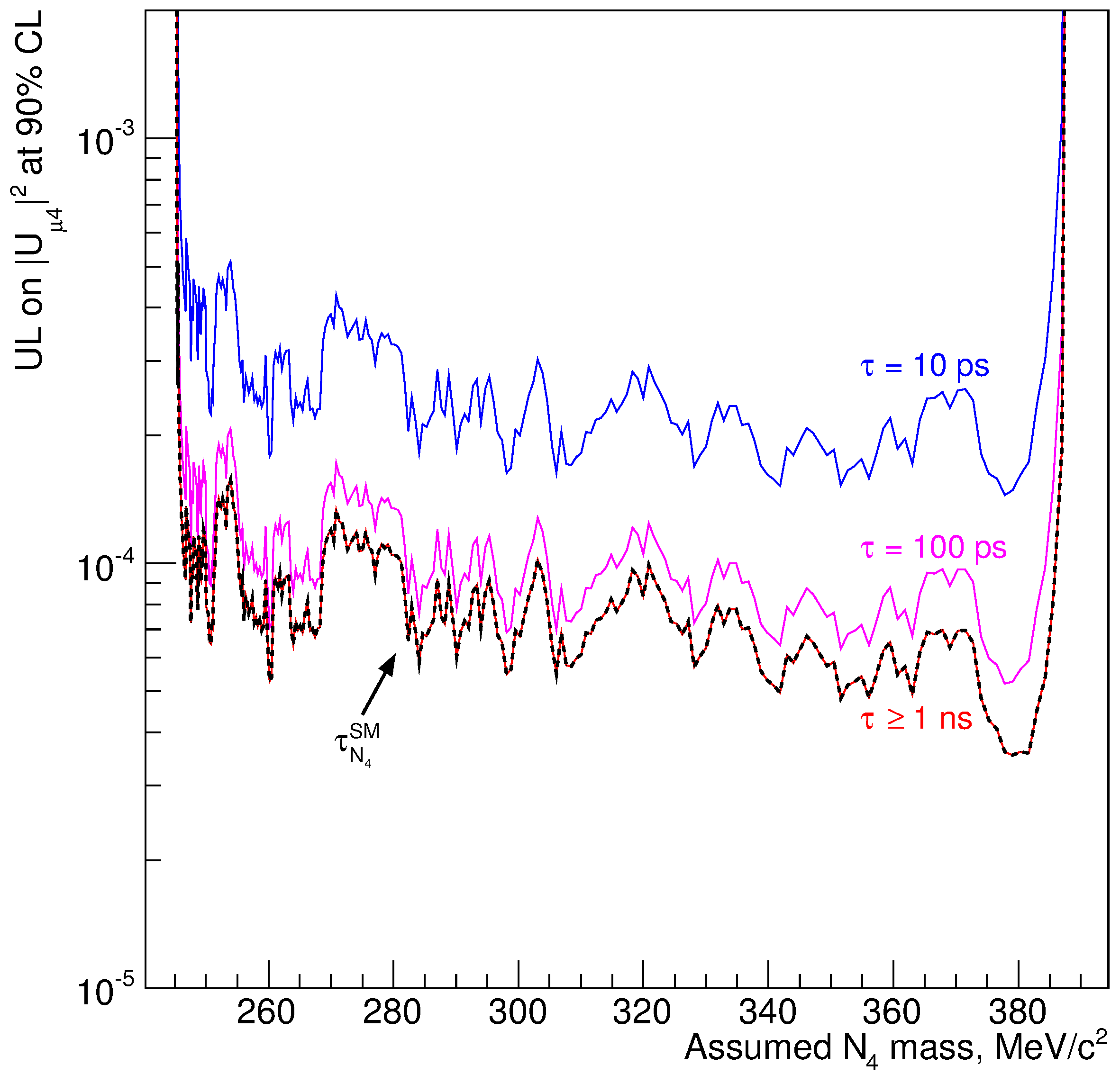}   
\end{center}
\endminipage \hfill
\minipage{0.5\textwidth}
\begin{center}
\includegraphics[width=0.9\textwidth]{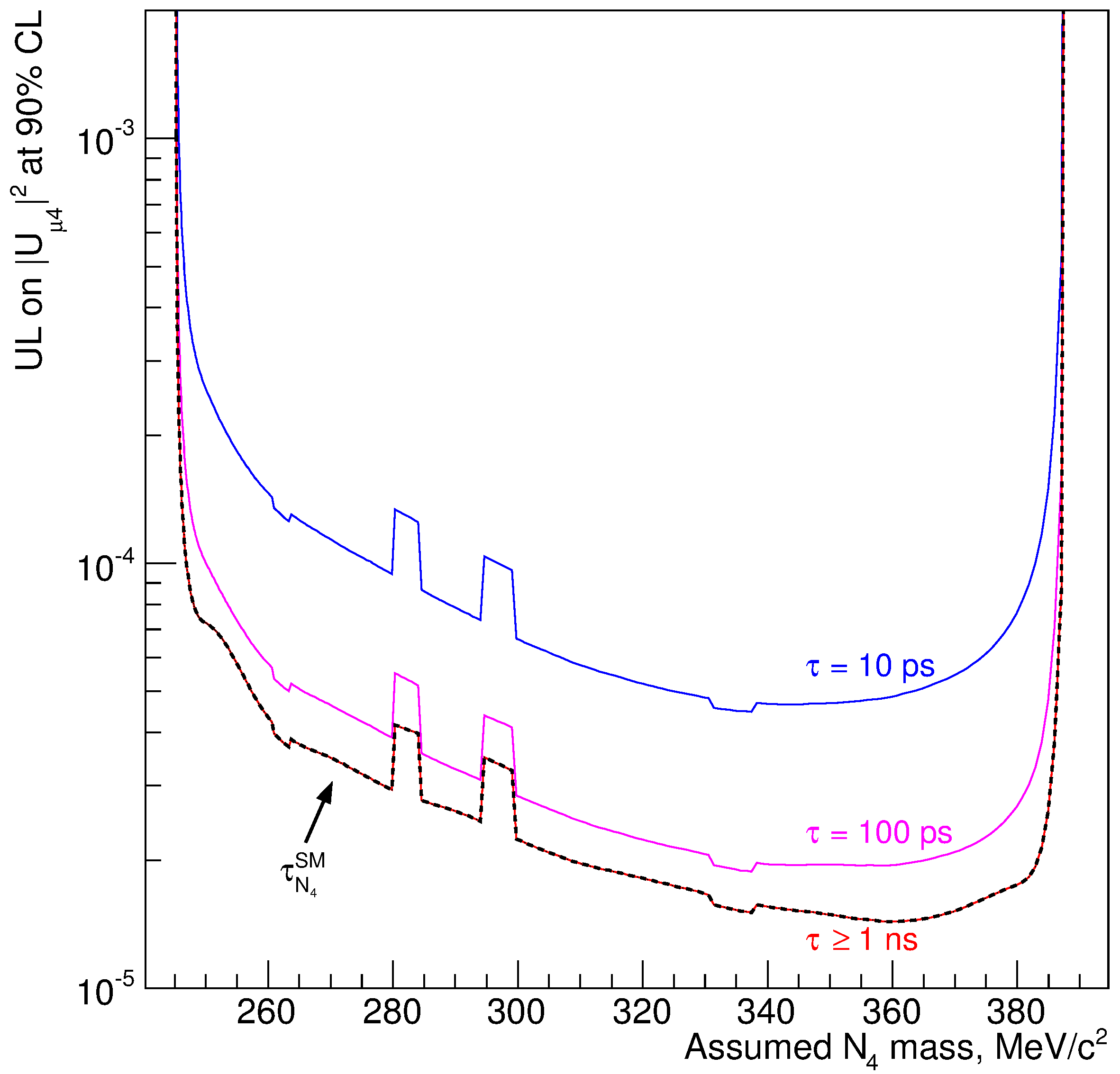}
\end{center}
\endminipage \hfill
\caption{Upper limit on the mixing parameter $|U_{4\mu}|$ as a function of the 
HNL mass obtained from the study of the $\pi\mu$ mass distribution in $K^{\pm}\to\pi^{\pm}\mu^{\pm}\mu^{\mp}$
events (left) and from the search for LNV decay $K^{\pm}\to\pi^{\mp}\mu^{\pm}\mu^{\pm}$ (right).}
\label{fig:na48-pimm-limits}
\end{figure}
The $K^{\pm}\to\pi^{\pm}\mu^{\pm}\mu^{\mp}$ and $K^{\pm}\to\pi^{\mp}\mu^{\pm}\mu^{\pm}$  events
were selected by requiring a three track vertex, one charged pion and two muons, same sign in case of $K^{\pm}\to\pi^{\mp}\mu^{\pm}\mu^{\pm}$  and
oposite sign in the case of $K^{\pm}\to\pi^{\pm}\mu^{\pm}\mu^{\mp}$.
This limits the search to short lived HNL.  
%, were required.
The signal was defined as $|M_{\pi\mu\mu} - M_{K^{\pm}}| < $ 8 MeV. 
A total of 3489 $K^{\pm}\to\pi^{\pm}\mu^{\pm}\mu^{\mp}$ candidates were selected with a 
background contribution of (0.36 $\pm$ 0.10)\%, while one event 
was observed in the $K^{\pm}\to\pi^{\mp}\mu^{\pm}\mu^{\pm}$ sample, with expected
background 1.16~$\pm$~0.87 events. 
The kaon flux was obtained from $K^{\pm}\to \pi^{\pm}\pi^+\pi^-$ events.
No significant signal was observed in the LNV mode, leading to the
%The lack of signal allowed to place an upper on the LNV decay, 
$Br( K^{\pm}\to\pi^{\mp}\mu^{\pm}\mu^{\pm}) < 8.6 \times 10^{-11}$, at 90\% confidence level.
The invariant mass distributions % $M_{\mu^+\mu^-}$, 
$M_{\pi^\pm \mu^\mp}$ in 
the LNC
%$K^{\pm}\to\pi^{\pm}\mu^{\pm}\mu^{\mp}$ 
sample and $M_{\pi^\mp \mu^\pm}$ 
in the LNV
%$K^{\pm}\to\pi^{\mp}\mu^{\pm}\mu^{\pm}$ 
sample was compared with the
Monte Carlo simulation. 
No signal consistent with the existence of a resonance was observed. 
This allowed to put an upper limit on the branching fractions $Br(K^{\pm}\to\mu^{\pm}N_4)\times Br(N_4\to\pi^{\pm}\mu^{\mp})$, $Br(K^{\pm}\to\mu^{\pm}N_4)\times Br(N_4\to\pi^{\mp}\mu^{\pm})$ and on the HNL 
mixing parameter $|U_{\mu4}|^2$, 
 %, and $Br(K^{\pm}\to\pi^{\pm}X)\times Br(X\to\mu^{\mp}\mu^{\pm})$, 
as shown in fig. \ref{fig:na48-pimm-limits} \cite{bib:NA48-lnv}.

\begin{figure}[htb]
\centering
\includegraphics[width=0.9\textwidth]{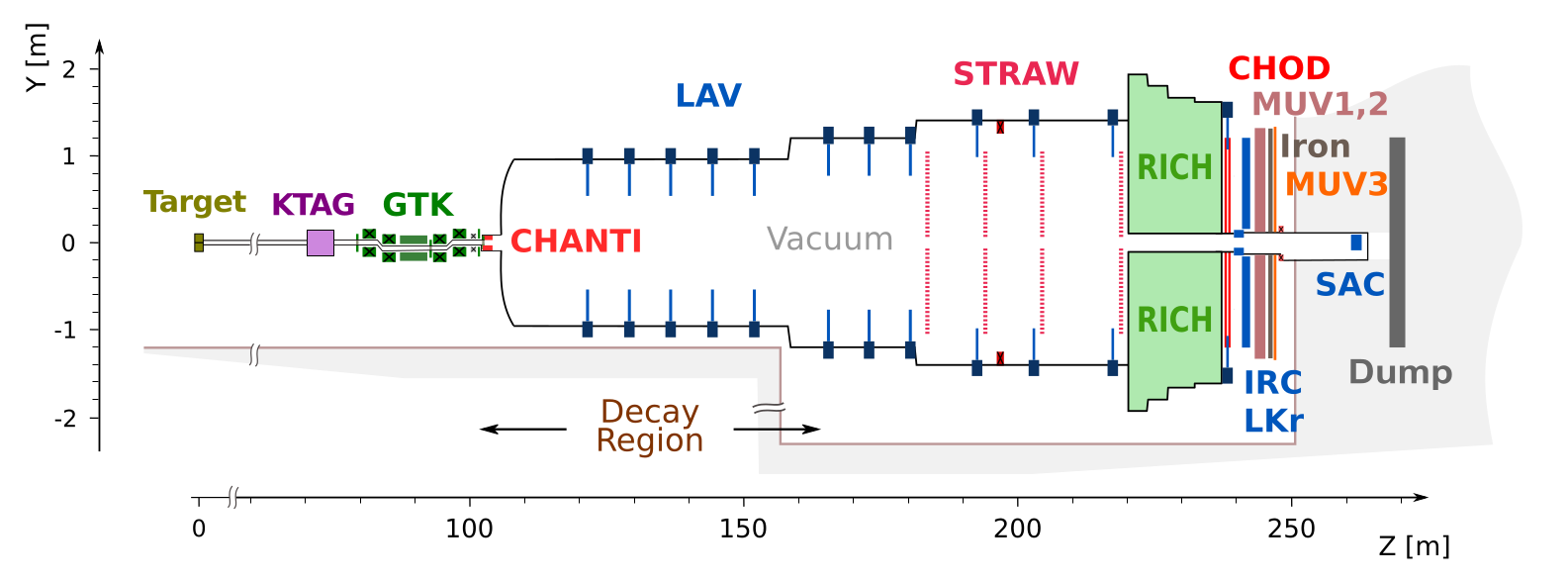}
\caption{Schematics of the NA62 experiment at CERN SPS.}
\label{fig:na62}
\end{figure}

The NA62 experiment \cite{bib:na62-exp}, schematically shown in fig. \ref{fig:na62}, is primarily devoted to the 
measurement of the $Br(K^+\to \pi^+\nu\bar{\nu})$ with 
10 \% precision \cite{bib:na62-pinn-2016}.
%The experimental setup, shown in fig. \ref{fig:na62} 
The primary proton beam from SPS interacts on a Be target
to produce high intensity 
positive beam of momentum 75~GeV/c~$\pm$~1\%, with 6\% $K^+$ content. 
%%%%%%%%%%%%%%%%%%%%%%%%%%%%%%%%%%%%%%%%%%%%%%%%%%%%%%%%%%%%%%%%%%%%%%%%%
%%
%%   use this format to include an .eps figure into your paper
%%
%%%%%%%%%%%%%%%%%%%%%%%%%%%%%%%%%%%%%%%%%%%%%%%%%%%%%%%%%%%%%%%%%%%%%%%%%%%
The secondary beam passes through a 
nitrogen filled 
threshold Cherenkov counter which  is used 
for positive kaon identification. 
% in the beam at a rate of 45 MHz.
%Time resolution of 100 ps was achieved which is important for the suppression of accidental
%background. 
%{\bf Gigatracker:} 
Three stations of thin silicon pixel detectors provide a measurement of the kaon momentum,
flight direction, and time. 
%The expected resolutions on the measured quantities will be $\sigma(p_K)/p_K \sim 0.2\%$ on momentum, 
%16 $\mu rad$ angular and  time resolution of 200 ps per station. 
A set of scintillating counters, CHANTI, provide a veto against interactions of the beam particles. 
The beam enters a 75 m long evacuated fiducial volume,
followed by a spectrometer with  a minimal material budget. 
It is made of four chambers of straw tubes operated in vacuum and
separated by the MNP33 dipole magnet. 
%They are  in order to provide
%momentum resolution of $\sigma(p)/p = (0.3 \oplus 0.008~p$ [GeV/c]))\%  with a minimal material budget. 
%{\bf LAV:}
A fast plastic scintillator charged hodoscope is used in the trigger. 
The NA48 liquid krypton calorimeter with renewed readout electronics 
measures the energy deposited by the particles and also 
serves as a 
photon veto for photons with angles from 1.5 to 8.5 mrad with inefficiency less than 
$10^{-5}$ for photons with energy above 10 GeV. 
Twelve rings of lead glass counters surrounding the decay region 
act as photon vetos 
for angles of the photons higher than 8.5 mrad with respect to the kaon flight direction. 
They are accompanied by two shashlyk type detectors covering photon angles down to zero.
The $\pi / \mu$ separation is based on the information from 
a neon filled ring imaging Cherenkov detector, measuring the 
velocity of the charged particles, and 
three stations of muon detectors.
Both KTAG and Gigatracker are exposed to the full 750 MHz hadron 
beam while the particle rate seen by the downstream detectors is at most 10 MHz.
The high kaon flux combined with 
precise kinematics measurement, particle identification, and hermeticity 
make the NA62 detector extremely powerful for the study of rare processes with kaons. 

\begin{figure}[hptb]
\minipage{0.5\textwidth}
\begin{center}
\includegraphics[width=0.9\textwidth]{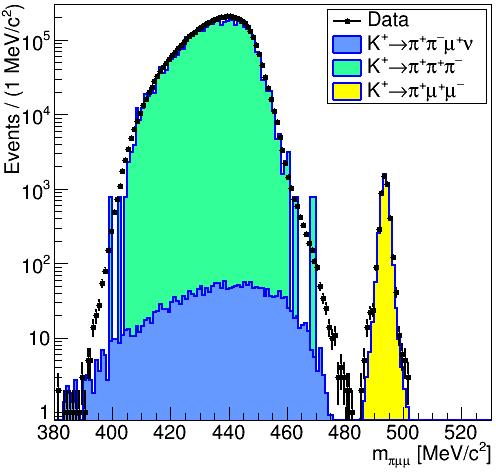}   
\end{center}
\endminipage \hfill
\minipage{0.5\textwidth}
\begin{center}
\includegraphics[width=0.9\textwidth]{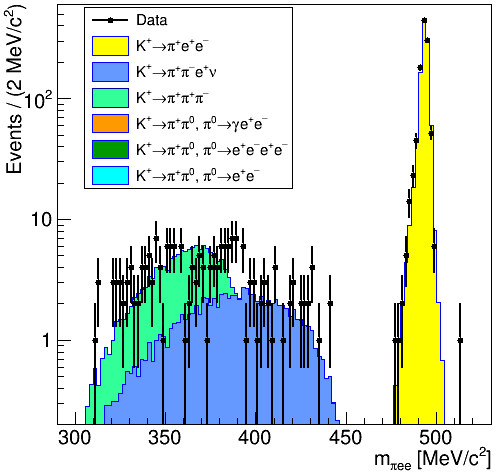}
\end{center}
\endminipage \hfill
\caption{Invariant mass spectrum for the reconstructed 
$\pi^{+}\ell^{+}\ell^{-}$ candidates for
the muon (left) and electron (right) channels. Only a partial NA62 dataset was used.}
\label{fig:na62-pill-prospects}
\end{figure}

Using a partial dataset, 
a search for the LNV modes $K^{+}\to\pi^{-}\ell^{+}\ell^{+}$, 
both for $\ell = \mu, e$,
was performed within NA62. 
%The search for HNL within NA62, with $N \to \pi^{\pm}\ell^{\mp}$, 
%was performed by studying the LNC modes $K^{+}\to\pi^{+}\ell^{+}\ell^{-}$ and 
%the LNV modes $K^{+}\to\pi^{-}\ell^{+}\ell^{+}$, both for $\ell = \mu, e$.
%A partial dataset was analysed and 
The obtained 
invariant mass spectrum 
for the LNC modes $K^{+}\to\pi^{+}\ell^{+}\ell^{-}$, 
which are used for normalization,
is shown in fig. \ref{fig:na62-pill-prospects}. 
The $K^{+}\to\pi^{+}\mu^{+}\mu^{-}$ sample is the world largest one, 
while the $K^{+}\to\pi^{+}e^{+}e^{-}$ analysis allowed  
the first observation of the decay in the mass range $m_{ee} < 140$ MeV/c$^2$. 
The NA62 detector was fully operational during the 2017 and 2018 data taking
and major improvements in the exclusive search for HNL is expected. 
The single event sensitivity for the LNV modes is estimated to be
$SES \sim 10^{-10}$ both for the electron and muon mode due to 
the negligible expected background.

\section{Inclusive HNL searches}

Inclusive search for HNL can be performed by looking for 
``bumps'' in the missing mass spectrum of the
$K^{\pm} \to \ell^{\pm} X $ decays, 
$|m_{miss}|^2 = (P_{K} - P_{\ell})^2$, with the charged lepton being 
the only reconstructed particle in the final state. 

During the early stage of NA62, a large data sample
of $K^+$ decays was collected in 2007 with a minimum bias trigger 
devoted to the
 measurement of $R_K = \Gamma(Ke2)/\Gamma(K\mu2)$ and the test of the lepton universality \cite{bib:na62-rk}. 
The nominal kaon momentum was 75 GeV/c and the MNP33 
current was increased to provide a $p_T$ kick of 265 MeV/c, 
leadint to 
momentum resolution of $\sigma(p)/p $ = (0.048  $\oplus $ 0.009  $p$ [GeV/c])\%. 
The nominal $P_K$ was calculated from $K_{3\pi}$ events.
The search for peaks was performed in the muon channel, in the mass range 
300 MeV/c$^2 < m_{miss}  < 375$~MeV/c$^2$ with a step of 1 MeV/c$^2$.
The dominant background was from $K^+ \to \pi^0\mu^+\nu$ and $K^+\to\mu^+\nu_{\mu}(\gamma)$ 
decays and from the muon halo. 
No signal exceeding 3$\sigma$ above the expected background was observed.
The Rolke-Lopez method was applied and 
%and 
an upper limit on $Br(K^+\to\mu^+N)$ was obtained \cite{bib:na62-2007-hnl-mmiss}.

\begin{figure}[hptb]
\minipage{0.48\textwidth}
\begin{center}
\includegraphics[width=0.9\textwidth]{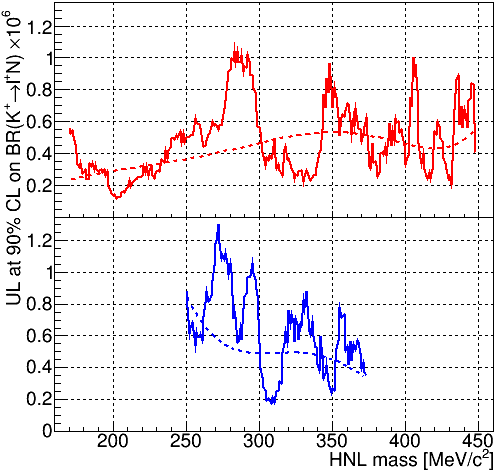}   
\end{center}
\endminipage \hfill
\minipage{0.48\textwidth}
\begin{center}
\includegraphics[width=0.9\textwidth]{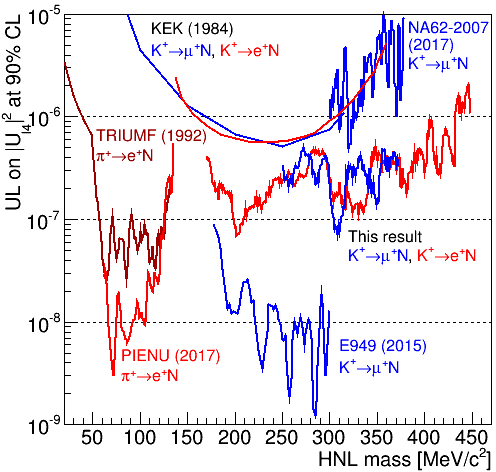}
\end{center}
\endminipage \hfill
\caption{Upper limits on the branching fraction for the $K^+\to\ell^+ N$ (left) and 
the obtained upper limits on the mixing parameter $|U_{\ell 4}|^2$ (right) as a 
function of the heavy neutrino mass. Data from previous experiments is also shown for comparison.}
\label{fig:na62-2015-hnl-mmiss}
\end{figure}

Using 5 days of data, recorded in 2015 with beam intensity
corresponding to 1\% of the nominal and a minimum bias trigger, the NA62 experiment
performed a search for HNL in the missing mass spectrum both for 
the electron and the muon mode \cite{bib:na62-2015-hnl-mmiss}. 
The already developed technique for the analysis of 2007 data was applied. 
The charged lepton momentum had to be between 5 GeV/c and 70 GeV/c. 
The search region for HNL was chosen to be 
170 (250) MeV/c$^2 < m_{miss} < $448 (373) MeV/c$^2$ for the 
electron (muon) channel. 
MC simulation was used to obtain the resolution $\sigma(m_N)$ 
and to calculate the acceptance for $K^+\to\ell^+ N$ events as a function 
of the HNL mass. 
The maximum value of the local signal significance $z = (N_{obs} - N_{exp})/\sqrt{N_{obs} + \delta N_{exp}^2}$ 
was 2.2, obtained for $m_N = 283$~MeV/c$^2$ in the electron channel.
The normalization was based on the 
reconstruction of the corresponding $K^+\to\ell^+ \nu_{\ell}$ decay 
and was used to obtain an upper limit 
on the branching fraction for the $K^+\to\ell^+ N$ decays, shown in fig. \ref{fig:na62-2015-hnl-mmiss}-left.
These upper limits on BR were translated into upper limits on the mixing parameters
$|U_{e4}|^2$ and $|U_{\mu4}|^2$, shown in fig. \ref{fig:na62-2015-hnl-mmiss}-right. 

As in the exclusive search case, a major improvements on the presented results could be expected 
with the present NA62 data.

\section{Future facilities}

The sensitivity of the presented searches at CERN SPS is 
limited by the statistics of the produced 
mesons, which afterwards may decay to final states with HNLs. 
This limitation could be overcome by entirely absorbing the primary proton beam 
and most of its interaction products in a
thick target, a technique known as beam-dump. 
The outgoing beam from the target consists of 
long lived neutral particles which enter a decay region. 

The number of the expected HNL in the detector
is a product of the produced HNLs and the probability 
for their observation \cite{bib:SHiP-HNL}.
The possible source of HNLs are the (semi)leptonic decays
of $\pi$, $K$, $D$, and $B$ mesons and the produced 
number of HNLs depends on the production of a given quark flavour in the target, 
its probability to hadronize to a certain meson and to decay to final states with HNL.
The detection probability is given by
\begin{equation}
P_{det} = \left[ e^{-\frac{l_{ini}}{{\gamma c \tau}}} -  e^{-\frac{l_{d}}{{\gamma c \tau}}}  \right] 
\times BR (N \to visible) \times \epsilon_{det}
\end{equation}
and depends on the length of the setup $l_{ini}$ before 
 the decay region, the length of the decay region $l_{d}$, 
 the lifetime $\tau$ of the HNL, the visible branching fraction and the 
 corresponding acceptance $\epsilon_{det}$.

%Since the kinematical properties of the HNL production state are unknown 
%the scenario that could be probed with the beam dump technique 
%is the HNL decays into visible states. 
%,  i.e. the beam-dump mode is necessarily an Exclusive search. 

The description of two selected facilities, 
a short term one (within 5 years) and a long term (10-15 years) one, 
discussed also within the Physics Beyond Coliders initiative \cite{bib:pbc}, follows.

\subsection{NA62 in beam dump mode}

The NA62 beryllium target is followed by two 
$\sim$11 nuclear interaction length 
water cooled copper-steel collimators (TAX) to stop the residual 
proton beam from SPS.
The dump mode operation could be provided by closing the first 
TAX. This allows to exploit the 
particle identification, tracking, and hermeticity of 
NA62 to search for long lived neutral particles, including HNL. 
The goal is to collect O($10^{18}$) POT in RUN3 \cite{bib:pbc-tommaso}. 

\begin{figure}[hptb]
\minipage{0.5\textwidth}
\begin{center}
\includegraphics[width=\textwidth]{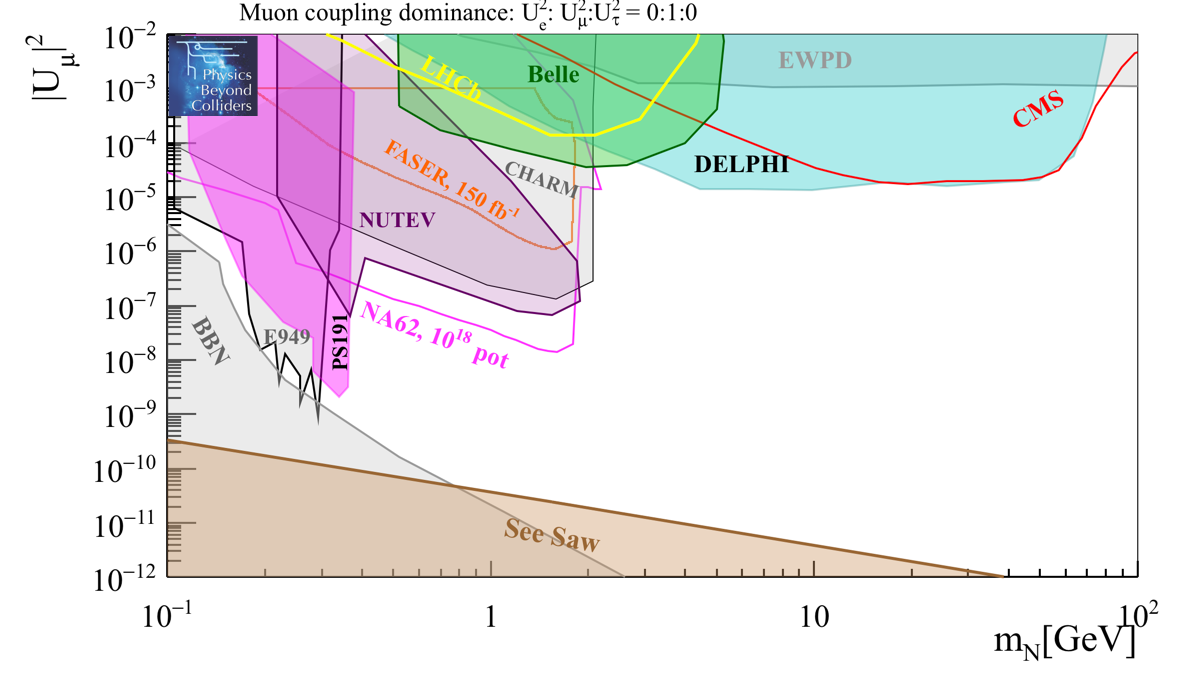}   
\end{center}
\endminipage \hfill
\minipage{0.45\textwidth}
\begin{center}
\includegraphics[width=\textwidth]{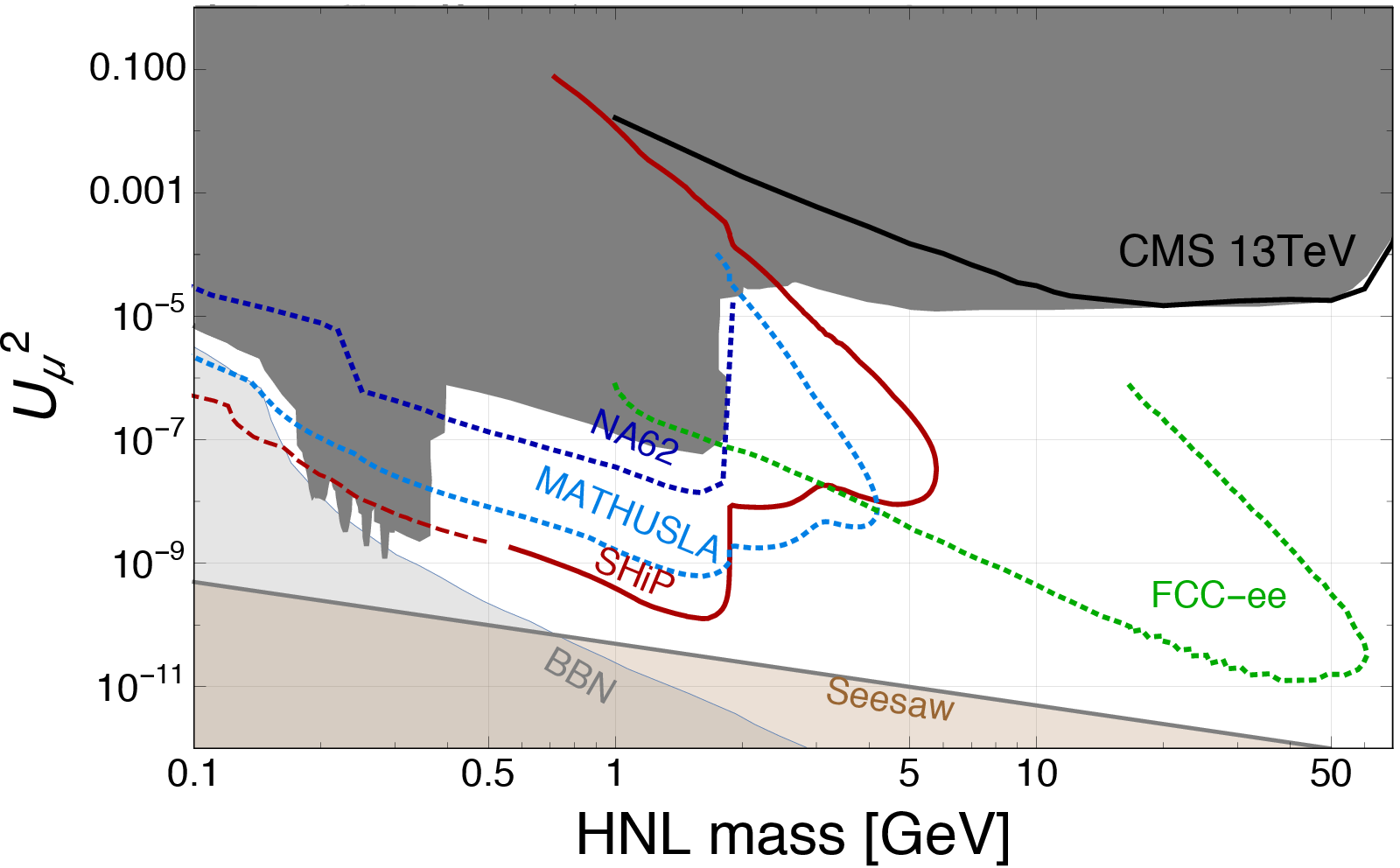}
\end{center}
\endminipage \hfill
\caption{Expected sensitivity of NA62 in beam dump mode 
(left) \cite{bib:pbc} and SHiP (right) \cite{bib:SHiP-HNL}
to the mixing parameter $|U_{\mu}|^2$ as a function of the HNL mass. 
Filled area is excluded by theory or previous experiments.}
\label{fig:na62dump-ship-hnl}
\end{figure}

The projected sensitivity of NA62 in beam dump mode (90\% CL upper limit)
for the case of dominant mixing with the second generation ($|U_e|^2: |U_\mu|^2:|U_\tau|^2 = 0:1:0$)
is shown in fig. \ref{fig:na62dump-ship-hnl}-left. 
It was obtained assuming zero background and 
the possibility to detect all two-track final states. The geometrical acceptance and the trigger 
efficiency were taken into account.

About $3\times 10^{16}$ POT have been collected by NA62 in a dedicated beam dump mode.
No background was achieved for fully reconstructed final states. Additional samples
collected in the presence of kaon beam are currently being used for dimuon background studies.

\subsection{SHiP}
\begin{figure}[htb]
\centering
\includegraphics[width=0.8\textwidth]{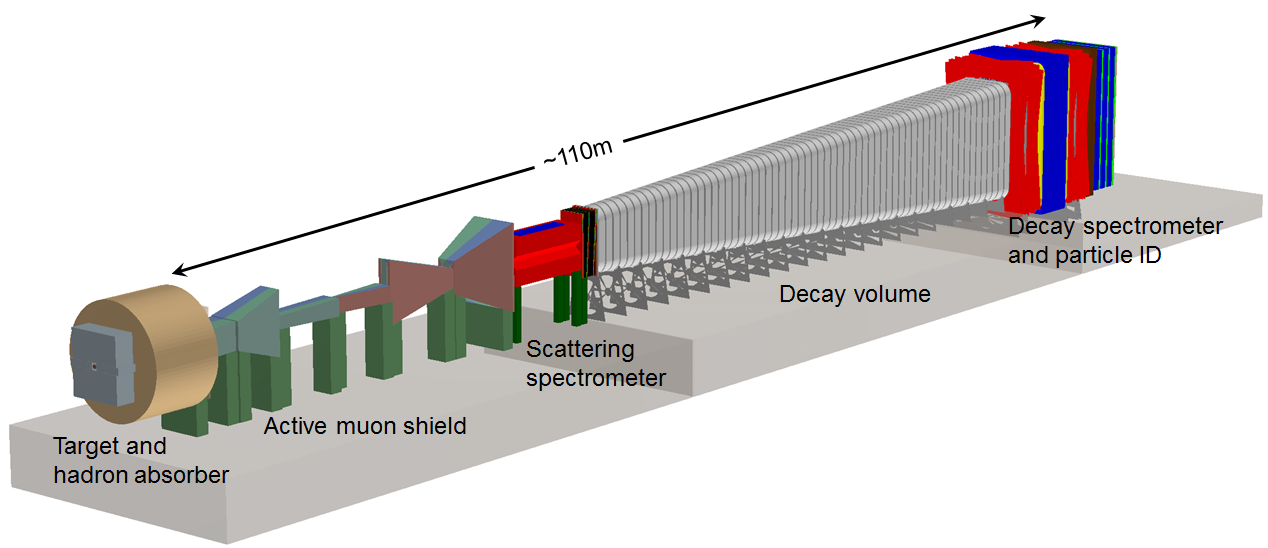}
\caption{SHiP experimental setup.}
\label{fig:ship}
\end{figure}

The Search for Hidden Particles (SHiP) experiment, shown schematically in fig. \ref{fig:ship}, 
aims to collect $2\times 10^{20}$ protons
on target in 5 years of data taking. 
It exploits a general purpose beam dump facility to complement the 
existing LHC programme in the search for New physics. 

The primary proton beam from SPS interacts in a 
very dense target (11 $\lambda_I$) 
followed by an hadron absorber. This assures
abundant production of heavy flavour states and also effectively 
stops the secondary protons and kaons. 
The target region is followed by an active muon shield to deflect the muons. 
The neutron interactions are reduced by evacuating the 60 m long decay region. 
The decay products are measured by a downstream magnetic spectrometer with 
large acceptance. A particle identification system is also foreseen.

The estimated sensitivity (90 \% confidence region) to HNL are obtained under the 
zero background assumption. 
The production fraction of $B_c$ mesons is unknown at SPS energies \cite{bib:gev-hnl} and 
is part of the systematic uncertainty. 
For the case of mixing with the second generation only the 
SHiP sensitivity is shown in fig. \ref{fig:na62dump-ship-hnl}-right. 
A region up to $m_N \sim$ O(6 GeV/c$^2$) and $|U_{\mu}|^2$ down to $10^{-10}$ 
could be covered, complementing the possible reach with a Future Circular Colider (FCC) in $e^+e^-$ mode.

The activities on the design and the construction of the detectors for the SHiP experiment are ongoing 
and almost every component has achieved at least first stage of prototyping \cite{bib:ship-status}.

\section{Conclusions}

A diverse heavy neutral leptons search programme 
is being executed at CERN with the NA62 experiment, 
covering both exclusive and inclusive events reconstruction. 
So far no signature of new states have been observed in
the (semi)leptonic decays of $K^{\pm}$. 
The obtained results are improving the existing limits
on the HNL mixing parameters in the considered mass range. 

A possible increase of the sensitivity is related to the 
application of the beam dump technique. 
Beyond the $K^+\to\pi\nu\bar{\nu}$ phase of NA62 experiment, NA62++ 
could collect $10^{18}$ POT in few months of 
operation during RUN3 (2021-2023). 
A dedicated beam dump facility with the SHiP experiment 
in the SPS North Area could 
further increase the statistics to $2\times10^{20}$ POT in 
5 years of operation, starting during RUN4 (after 2027). 

Both NA62++ and SHiP are part of the PBC working group and 
provide input to the European Strategy for Particle Physics.

%\Acknowledgements
%VK acknoledges support from LNF-INFN
%under agreement 
%between University of Sofia and LNF 
%SU-LNF 70-06-497/07-10-2014.

\end{document}